\documentclass[twocolumn,prl,showpacs,aps]{revtex4}
\usepackage{amsmath}
\usepackage{graphicx}
\usepackage{amssymb}
\usepackage{esint}
\usepackage{color,soul}

\makeatletter
\@ifundefined{textcolor}{}
{%
 \definecolor{BLACK}{gray}{0}
 \definecolor{WHITE}{gray}{1}
 \definecolor{RED}{rgb}{1,0,0}
 \definecolor{GREEN}{rgb}{0,1,0}
 \definecolor{BLUE}{rgb}{0,0,1}
 \definecolor{CYAN}{cmyk}{1,0,0,0}
 \definecolor{MAGENTA}{cmyk}{0,1,0,0}
 \definecolor{YELLOW}{cmyk}{0,0,1,0}
 }

\makeatletter

\usepackage{dcolumn}% Align table columns on decimal point
\usepackage{bm}% bold math
\usepackage{amsthm}\@ifundefined{definecolor}
 {\@ifundefined{definecolor}
 {\usepackage{color}}{}
}{}
\usepackage{multirow}

\begin{document}
\newtheorem{conjecture}{Conjecture}\newtheorem{corollary}{Corollary}\newtheorem{theorem}{Theorem}
\newtheorem{lemma}{Lemma}
\newtheorem{observation}{Observation}
\newtheorem{definition}{Definition}\newtheorem{remark}{Remark}\global\long\global\long\def\ket#1{|#1 \rangle}
 \global\long\global\long\def\bra#1{\langle#1|}
 \global\long\global\long\def\proj#1{\ket{#1}\bra{#1}}

\title{Is Bell inequality violated once measurement independence is relaxed?}

\author{Minsu Kim$^{1}$}
\email{shuypigel@gmail.com}
\author{Jinhyoung Lee$^{2}$}
\email{hyoung@hanyang.ac.kr}
\author{Sang Wook Kim$^{3}$}
\email{swkim0412@pusan.ac.kr}

 \affiliation{$^{1}$Department of Physics, Pusan National University, Busan 609-735, Korea \\
$^{2}$ Quantum Information Laboratory, Department of Physics, Hanyang University, Seoul 133-791, Korea \\
$^{3}$Department of Physics Education, Pusan National University, Busan 609-735, Korea }

\date{\today}

\pacs{03.65.Ta, 03.65.Ud, 03.67.-a, 03.65.Ca}
\begin{abstract}
It has been believed that statistical inequality such as Bell inequality should be modified once measurement independence (MI), the assumption that observers can freely choose measurement settings without changing the probability distribution of hidden variables, is relaxed. However, we show that there exists the possibility that Bell inequality is still valid even if MI is relaxed. MI is only a sufficient condition to derive Bell inequality when both determinism and setting independence, usually called as local realism, are satisfied. We thus propose a new condition necessary and sufficient for deriving Bell inequality, called as concealed measurement dependence.
\end{abstract}
\maketitle

%\section{Introduction}

One of the most remarkable achievement of modern physics is that quantum mechanics violates certain statistical inequalities such as Bell inequality \cite{Bell64,Bell66,Clauser74, Rowe01}. Bell inequality is derived based upon several physical postulates; namely determinism, setting independence and measurement independence (MI). Determinism is the property that an outcome of any physical observable has a definite value all the time. Setting independence implies that the probability of observing an event associated with one setting is independent of the other setting so that it prohibits any information from transmitting faster than light, called as no-signaling.

MI is the assumption that measurement settings can be chosen independently of any underlying variables describing a system. Compared with determinism and setting independence, MI has not been seriously considered in literatures since it is believed that experimenters can freely choose an experimental setup. In this regard MI is often associated with the so-called ``free will'' of experimenters \cite{Hall11}. As clearly discussed in Ref.\cite{Brans88}, however, there are {\em no} free or random events in a hidden variable theory based upon truly classical mechanics so that MI cannot be naturally ensured. If determinism and setting independence are suspected, fair attention should be paid to MI.

So far focus has lied at constructing singlet correlation, maximally violating Bell inequality, by relaxing MI. Brans has firstly shown that  singlet correlation can be reproduced by completely relaxing MI \cite{Brans88}. More quantitative studies to obtain singlet correlation by partially relaxing MI has been performed by introducing some measures \cite{Kofler06, Hall10, Barrett11} or by using models \cite{Hall11, Lorenzo12, Banik12, Zela13}. In fact, singlet correlation has also been acquired by relaxing no-signaling \cite{Toner03, Pawlowski10}, determinism \cite{Branciard08,Lorenzo12a,Hall10a, Kar11}, or both MI and outcome independence \cite{Groeblacher07, Cerf05, Brunner08, Lorenzo13}.

In this paper, instead of constructing singlet correlation by relaxing MI, we focus on a question; whether MI is sufficient, necessary or both necessary and sufficient condition to fulfill Bell inequality when all the other conditions such as determinism, setting independence, and so on are assumed. We show that Bell inequality can be still valid even if MI is relaxed. MI is only a sufficient condition to derive Bell inequality when determinism and setting independence are satisfied. We thus propose a new condition necessary and sufficient for deriving Bell inequality, called as concealed measurement dependence (CMD). We also find that our CMD hidden variable model may violate no-signaling even if locality is assumed.

We consider Bell inequalities of correlations in the form of
\begin{equation}
\label{eq: Bell inequality}
\sum_{i,j=1,2} w_{ij} E(A_i, B_j) \le C
\end{equation}
where $C$ is a classical upper bound and $w_{ij}$ are weight coefficients of real numbers. Here, $E(A,B)$ is the average of spin correlation when Alice and Bob measure their spins of the correlated particles along the direction $A$ and $B$, respectively, assuming that they are space-like separated. For deterministic hidden variable models, Bell inequalities~(\ref{eq: Bell inequality}) become CHSH ones with $C=2$ when $w_{11}=w_{12}=w_{21}=-w_{22}= \pm 1$ \cite{CHSH69}.

An average of any correlation is in the most general form expressed as
\begin{eqnarray}
E(X,Y)&&=\sum_{\alpha,\beta= \pm 1} \alpha \beta \, P_{XY}(\alpha,\beta),
\label{eq: E general form}
\end{eqnarray}
where $P_{XY}(\alpha,\beta)$ is joint probability of obtaining the outcomes $\alpha$ and $\beta$ when Alice measures $X$ (direction) and Bob does $Y$ (direction), respectively. Here, we assume that systems are spin-1/2 particles and each spin component is dichotomic, i.e., $\alpha$ and $\beta$ are $\pm 1$. Hidden variable theories assume that the joint probability function is given by
\begin{eqnarray}
P_{XY}(\alpha,\beta) = \sum_{\lambda} P_{XY}(\alpha,\beta|\lambda) P_{XY}(\lambda)
\label{eq: E general form1}
\end{eqnarray}
where $\lambda$ denotes hidden (or underlying) variables. The summation $\sum_\lambda$ is replaced by integration when the underlying variable is continuous. One may move measurement dependence on $X$ and $Y$ to conditions:
\begin{eqnarray}
P(\alpha,\beta|X,Y) = \sum_{\lambda} P(\alpha,\beta|\lambda,X,Y) P(\lambda|X,Y).
\label{eq: E general form2}
\end{eqnarray}

Locality assumes that physical properties can not be influenced by space-like separated events in a superluminal way. This implies the joint probabilities conditioned by hidden-variable is factorable:
\begin{equation}
P(\alpha,\beta|\lambda,X,Y) = P(\alpha|\lambda,X) P(\beta|\lambda,Y).
\label{eq:locality}
\end{equation}
This assumption involves two types of independences; (measurement) setting independence and outcome independence. More explicitly, based on Bayesian rule, $P(\alpha,\beta|\lambda,X,Y) = P(\alpha|\beta,\lambda,X,Y) P(\beta|\lambda,X,Y)$, setting independence implies that instantaneous change of one setting does not affect the probability distribution in another setting separated (possibly) with space-like distance from the one: $P(\beta|\lambda,X,Y)  = P(\beta|\lambda,X',Y) \equiv P(\beta|\lambda,Y)$ and similarly $P(\alpha|\beta,\lambda,X,Y)= P(\alpha|\beta,\lambda,X,Y') \equiv P(\alpha|\beta,\lambda,X)$. Furthermore, outcome independence implies that obtaining a particular outcome in one measurement does not affect the probability distribution in another measurement: $P(\alpha|\beta,\lambda,X) = P(\alpha|\lambda,X)$. This is usually assumed to be correct once determinism is fulfilled \cite{Hall11}.

MI is the assumption of influence of the choices of measurement settings on the probability distribution of hidden variables, $P(\lambda|X,Y)$. It says that Alice and Bob can freely choose their measurement settings, leaving the hidden variable distribution $P(\lambda|X,Y)$ intact. In other words,
\begin{equation}
P(\lambda|X,Y)=P(\lambda|X^\prime,Y^\prime)\equiv P(\lambda)
\label{eq:MI}
\end{equation}
for $X' \ne X$ or $Y' \ne Y$.

Fine proved that the existence of a hidden variable model which assumes determinism, locality and MI is a necessary and sufficient condition that Clauser-Horne inequalities hold \cite{Fine82,Clauser74}, which we call a MI deterministic hidden variable model. Clauser-Horne inequalities are equivalent to CHSH ones with perfectly efficient measurements \cite{Clauser74}. If any one of the assumptions is relaxed, on the other hand, Bell inequality will be altered such that the absolute value of a classical bound increases. It thus becomes more difficult to violate it in quantum mechanics. The increase of classical bound, however, does not always occur, even when MI is relaxed. Interestingly there exists certain situation where Bell inequality remains intact. We shall find such conditions, called as concealed measurement dependence (CMD), which is the main theme of this paper. A local deterministic hidden variable model with the MI relaxed by CMD is referred to as a CMD model.

CMD is defined as
\begin{eqnarray}
 \label{eq:CMD2}
E(X,Y) &=& E_{A_1B_1}(X,Y)
\end{eqnarray}
with all possible settings $X \in \{A_1,A_2\}$ and $Y \in \{B_1,B_2\}$. Note that $A_1$ and $B_1$ represented as the subscript of the righthand side simply play roles of references; One can choose any other settings instead of $A_1$ and $B_1$.
Here, we introduce the generalized correlations defined as \footnote{We note that $E_{X'Y'}$ should not be regarded as any physical quantities. They should instead be understood as mathematical objects in describing the criteria of Bell inequalities.}
\begin{eqnarray*}
&& E_{X'Y'}(X,Y) = \sum_\lambda \alpha(\lambda,X)\beta(\lambda,Y) P(\lambda|X',Y').
\end{eqnarray*}
CMD does not automatically guarantee MI since even with $P(\lambda|X,Y) \ne P(\lambda|X',Y')$ Eq.~(\ref{eq:CMD2}) can be satisfied. By every CMD model, nevertheless, Bell inequalities of Eq.~(\ref{eq: Bell inequality}) remain intact,
\begin{eqnarray}
\label{eq: Bell by CMD2}
\sum_{i,j=1,2} w_{ij} E(A_i,B_j) = \sum_{i,j=1,2} w_{ij} E_{A_1B_1}(A_i,B_j) \le C,
\end{eqnarray}
where the classical upper bound is equal to that of MI deterministic hidden variable model. This directly results from Eq.~(\ref{eq:CMD2}). CMD is weaker than MI since it is the restrictions on correlations rather than on the distributions of hidden variables. It is worth noting that the existence of CMD does not contradict Fine's proof \cite{Fine82} which requires the correlations of CMD model to be simulated by certain MI deterministic hidden variable model. Here, this is the case since $E_{A_1B_1}(X,Y)$ are averaged over {\em single} hidden-variable distribution $P(\lambda|A_1,B_1)$ like usual MI deterministic hidden variable models.

We discuss a subtle issue on CMD. Even though locality is assumed, CMD models may violate no-signaling expressed as
\begin{equation}
P(\alpha|X,Y) = P(\alpha|X,Y') ~{\rm and}~ P(\beta|X,Y) = P(\beta|X',Y)
\label{eq:no_signaling}
\end{equation}
with $P(\alpha(\beta)|X,Y)=\sum_{\beta(\alpha)} P(\alpha,\beta|X,Y)$. By using Eq.~(\ref{eq: E general form2}), the conditional probabilities of Eq.~(\ref{eq:no_signaling}) are rewritten as
\begin{equation}
P(\alpha(\beta)|X,Y) = \sum_{\beta(\alpha)} \sum_{\lambda} P(\alpha,\beta|\lambda,X,Y) P(\lambda|X,Y).
\end{equation}
Here locality, setting independence, and outcome independence are all associated with $P(\alpha,\beta|\lambda,X,Y)$, but have nothing to do with $P(\lambda|X,Y)$. Thus, $P(\alpha(\beta)|X,Y)$ may still depend on $X$ or $Y$ once the MI defined as Eq.~(\ref{eq:MI}) is relaxed; In the physical viewpoint signal can be transferred by altering measurement settings. We can also explain it in a slight different way. In CMD models MI is relaxed in a very specific form, Eq.~(\ref{eq:CMD2}); CMD includes restriction not on local expectations $E(X)$ and $E(Y)$ but only on the correlations $E(X,Y)$. Even with CMD, $E_{XY}(X) = E_{XY'}(X)$ and $E_{XY}(Y) = E_{X'Y}(Y)$ are not guaranteed, where $E_{XY}(Z) = \sum_{\lambda} \alpha(\lambda,Z) \, P(\lambda|X,Y)$. Therefore, no-signaling (\ref{eq:no_signaling}) may be violated since probabilities are directly related to their expectations, i.e. $P(\alpha|X,Y) = [1 + \alpha \, E_{XY}(X)]/2$ and $P(\beta|X,Y) = [1 + \beta \, E_{XY}(Y)]/2$ \footnote{One may also expand joint probabilities in terms of expectations by $P(\alpha,\beta|X,Y) = (1+ \alpha \, E_{XY}(X) + \beta \, E_{XY}(Y) + \alpha \beta \, E_{XY}(X,Y))/4$.}.
It is remarkable that {\em satisfying the Bell inequalities of MI deterministic hidden variable models is not sufficient for no-signaling in CMD models.}

Now we find the relationships among the sets of hidden variable models satisfying CMD, Bell inequalities, and no-signaling when MI is relaxed but locality and determinism is still assumed. First, CMD implies Bell inequalities by definition. We will show below that Bell inequalities also imply CMD, so that the sets of CMD is equivalent to those of Bell inequalities. Second, as discussed above, CMD does not guarantee no-signaling. However, MI model should exist if both Bell inequalities and no-signaling are satisfied due to Fine's proof \cite{Fine82}. It implies that not only there exists non-zero intersection between the sets of Bell inequalities and no-signaling, but it also should contain the sets representable by MI. The intersection is larger than the sets of MI deterministic hidden variable model since CMD model deals with the correlation of only two variables. The relations discussed here is schematically summarized in Fig.~1. We emphasize that {\em CMD models are not distinguishable from MI deterministic hidden variable models by only testing the Bell inequalities of the MI deterministic hidden variable models.}

\begin{figure}[!h]
 \centering \includegraphics[scale=0.3]{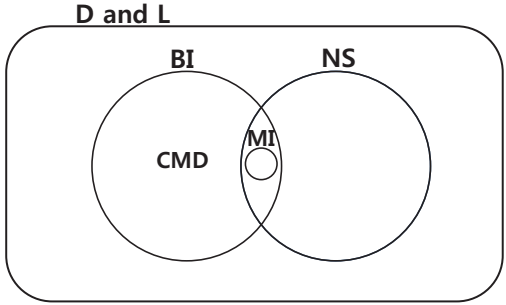} \caption{Schematic diagram for the sets of hidden variable distributions satisfying Bell inequalities (BI), no-signaling (NS), and CMD when determinism (D) and locality (L) are assumed but MI is relaxed. The `MI' denotes the set representable with MI deterministic hidden variable model.}
\label{fig:01}
\end{figure}

The CMD condition, Eq.~(\ref{eq:CMD2}), can be more explicitly expressed. We consider the local deterministic hidden variable model satisfying  CMD rather than MI. Without loss of generality measurement-dependent hidden-variable distributions can always be expressed as
\begin{equation}
\label{eq:xi_XY_lambda}
P(\lambda|X,Y) = P(\lambda|A_1,B_1) + \xi_{XY\lambda}
\end{equation}
as far as normalization conditions, $\sum_\lambda \xi_{XY\lambda} = 0$, and positivity $P(\lambda|X,Y) \ge 0$ are fulfilled. It simply means that the probability distributions are expressed as the difference from the reference distribution, namely here $P(\lambda|A_1,B_1)$. Therefore, $\xi_{A_1B_1\lambda} = 0$ is trivially satisfied. Putting Eq.~(\ref{eq:xi_XY_lambda}) into Eq.~(\ref{eq:CMD2}) the CMD conditions are written as
\begin{eqnarray}
\sum_{(X,Y) \ne (A_1,B_1)} \sum_{\lambda} M^{X'Y'\eta}_{XY\lambda} \xi_{XY\lambda} = 0,
\label{eq:CMD4}
\end{eqnarray}
where $M^{X'Y'\eta=1}_{XY\lambda} = \alpha(\lambda,X) \beta(\lambda,Y) \delta_{XX'} \delta_{YY'}$ for the CMD conditions ($\eta=1$) and $M^{X'Y'\eta=2}_{XY\lambda} = \delta_{XX'} \delta_{YY'}$ for the normalization conditions ($\eta=2$), or more compactly as
\begin{eqnarray}
\mathbf{M} \mathbf{\xi} = \mathbf{0}.
\end{eqnarray}
Here $\mathbf{\xi}$ is a vector of $48 (=3\times 2^4)$ dimensions, $\mathbf{0}$ a null vector, and $\mathbf{M}$ a $6 \times 48$ matrix. This implies that $\mathbf{\xi}$ belongs to the kernel of $\mathbf{M}$ denoted as $\ker(\mathbf{M})$. All six row vectors of $\mathbf{M}$ are mutually orthogonal so that the dimension of $\ker(\mathbf{M})$, or the nullity of $\mathbf{M}$ is 42. The conventional MI forms 0-dimensional kernel since it is achieved only by $\mathbf{\xi} = \mathbf{0}$. Even if MI is relaxed, i.e. $\mathbf{\xi} \ne \mathbf{0}$, Bell inequalities remain intact as long as $\vec{\xi} \in \ker(\mathbf{M})$ is satisfied. In a similar way, the no-signaling conditions can also be written as
\begin{eqnarray}
\mathbf{N} \mathbf{\xi} \equiv \sum_{(X,Y) \ne (A_1,B_1)}\sum_{\lambda} N^{j \eta}_{XY\lambda} \xi_{XY\lambda}=\mathbf{0},
\end{eqnarray}
where $N^{j\eta=1}_{XY\lambda} = \alpha(\lambda,X) \delta_{XA_j} \left( \delta_{YB_1} - \delta_{YB_2} \right)$ and $N^{j\eta=2}_{XY\lambda} = \beta(\lambda,Y) \left(\delta_{XA_1} - \delta_{XA_2} \right) \delta_{YB_j}$ for no-signaling, and $N^{j\eta=3}_{XY\lambda} = \delta_{XA_j} \delta_{YB_1}$ and $N^{j\eta=4}_{XY\lambda} = \delta_{XA_j} \delta_{YB_2}$ for normalization. $\mathbf{N}$ is a $7 \times 48$ rectangular matrix. All 7 row vectors of $\mathbf{N}$ are mutually orthogonal, similarly in the CMD matrix $\mathbf{M}$, so that $\ker(\mathbf{N})$ forms $41$-dimensional subspace.

Using Eq.~(\ref{eq:xi_XY_lambda}) Bell inequalities are rewritten as
\begin{eqnarray}
\label{eq: Bell by CMD2}
\sum_{i,j=1,2} w_{ij} E_{A_iB_j}(A_i,B_j) = C + \gamma,
\end{eqnarray}
with
\begin{eqnarray}
\gamma &=& \sum_{(i,j) \ne (1,1)} w_{ij} \sum_{(X,Y) \ne (A_1,B_1)} \sum_{\lambda} M^{A_iB_j \eta=1}_{XY\lambda} \xi_{XY\lambda} \nonumber \\
&\equiv& \mathbf{w} \cdot \mathbf{\tilde{\xi}}. \label{eq:xi_perp}
\end{eqnarray}
Here $\mathbf{w} = (w_{12},w_{21},w_{22})^T$, where $T$ denotes the transpose, and $\mathbf{\tilde{\xi}} = \mathbf{M}^{\eta=1} \mathbf{\xi} = \mathbf{M}^{\eta=1} \mathbf{\xi}^\perp$ with $\mathbf{\xi}^\perp= \mathbf{\xi} - \mathbf{\xi}^\parallel$ and $\mathbf{\xi}^\parallel \in \ker(\mathbf{M})$ according to $\mathbf{M}^{\eta=1} \mathbf{\xi}^\parallel =0$. Considering all possible $\mathbf{w}$, the maximum of $\gamma$ is given as
\begin{equation}
\gamma_M = \sup_{\mathbf{w}} \mathbf{w} \cdot \mathbf{M}^{\eta=1} \mathbf{\xi}^\perp.
\end{equation}
One can thus say that the increase of the classical bound is determined by $\mathbf{\xi}^\perp$. It is worth noting that the classical bound defined as $\min\{2+3{\cal M}, 4\}$ with
\begin{eqnarray*}
{\cal M} \equiv \sup_{X,X',Y,Y'} \sum_{\lambda} \left|\xi_{XY\lambda} - \xi_{X'Y'\lambda} \right|
\end{eqnarray*}
has been proposed in Ref.~\cite{Hall10}, but this differs from $\gamma_M$.

According to Eq.~(\ref{eq:xi_perp}) the fact that Bell inequalities are satisfied implies $\gamma=0$, or equivalently
\begin{equation}
\mathbf{w} \cdot \mathbf{M}^{\eta=1} \mathbf{\xi} = \mathbf{0}.
\label{eq:CMD converse}
\end{equation}
As far as every Bell inequalities represented by all possible $\mathbf{w}$'s are concerned, Eq.~(\ref{eq:CMD converse}) is fulfilled if and only if $\mathbf{M}^{\eta=1} \mathbf{\xi} = \mathbf{0}$. Together with the normalization condition $\mathbf{M}^{\eta=2} \mathbf{\xi} = \mathbf{0}$, which should be satisfied in any case, we reach $\mathbf{M} \mathbf{\xi} = \mathbf{0}$. It proves Bell inequalities imply CMD, the converse proposition of the statement that CMD implies satisfying Bell inequalities. Therefore, the sets of the hidden variable models satisfying CMD is equivalent to those of Bell inequalities.

Our results can be summarized by using the kernels of $\mathbf{M}$ and $\mathbf{N}$ as follows
\begin{itemize}
\item[(S1)] If and only if $\mathbf{\xi} \in \ker(\mathbf{M})$ and $\mathbf{\xi} \in \ker(\mathbf{N})$, there exists a MI deterministic hidden variable model.
\item[(S2)] If $\mathbf{\xi} \in \ker(\mathbf{M})$, there exists a CMD model. Furthermore, if $\mathbf{\xi} \in \ker(\mathbf{M})$ but $\mathbf{\xi} \notin \ker(\mathbf{N})$, no MI deterministic hidden variable model exists.
\item[(S3)] If $\mathbf{\xi} \notin \ker(\mathbf{M})$, there exists a measurement-dependent model satisfying Bell inequalities with the increased classical bound determined by $\mathbf{\xi}^\perp$.
\item[(S4)] If $\mathbf{\xi} \notin \ker(\mathbf{M})$ and $\mathbf{\xi} \in \ker(\mathbf{N})$, there exists a measurement-dependent and no-signaling model satisfying Bell inequalities with the increased classical bound determined similarly in (S3).
\end{itemize}

In conclusion, we have shown that Bell inequalities can be still valid even if MI is relaxed. The necessary and sufficient condition for satisfying Bell inequalities is given as CMD if both determinism and locality are assumed. We also find that our CMD models may violate the no-signaling condition even if locality is assumed. By using explicit probability distributions of measurement-dependent hidden variables we obtain the mathematical expressions satisfying CMD so that we find the correct mathematical formula of the classical bound.

%\section*{Acknowledgments}
This research was supported by Basic Science Research Program through the National Research Foundation of Korea(NRF) funded by the Ministry of Science, ICT and future Planning (2009-0087261, 2013R1A1A2011438, 2010-0018295, and 2010-0015059).

\bibliography{/Users/jlee/Documents/Research/Work/references}

\end{document}